\documentclass[aps,prl,twocolumn,showpacs,groupedaddress]{revtex4}

\usepackage{graphicx}
\usepackage{dcolumn}
\usepackage{bm}
\usepackage{amssymb}

\begin{document}


\title{Swinging and Tumbling of Fluid Vesicles in Shear Flow}

\author{Hiroshi Noguchi}
\email[]{e-mail: hi.noguchi@fz-juelich.de}
\author{Gerhard Gompper}
\affiliation{
Institut f\"ur Festk\"orperforschung, Forschungszentrum J\"ulich, 
52425 J\"ulich, Germany}

\date{\today}

\begin{abstract}
The dynamics of fluid vesicles in simple shear flow is studied
using mesoscale simulations of dynamically-triangulated surfaces, as well
as a theoretical approach based on two variables, a shape parameter and 
the inclination angle, which has no adjustable parameters. We show 
that between the well-known tank-treading and tumbling states,  
a new ``swinging'' state can appear. We predict the dynamic phase diagram 
as a function of the shear rate, the viscosities of the 
membrane and the internal fluid, and the reduced vesicle volume.
Our results agree well with recent experiments.
\end{abstract}
\pacs{87.16.Dg,83.50.-v,87.17.Aa}

\maketitle

The dynamics of deformable objects such as liquid droplets~\cite{rall84,bart85}, 
lipid vesicles~\cite{haas97,kant06,seif99,misb06,krau96,beau04,nogu04,nogu05}, 
red blood cells~\cite{fisc78,kell82,tran84,pozr03}, 
and elastic capsules~\cite{chan93,skot06,fink06} in flows has received 
increasing attention experimentally, theoretically, and numerically in recent 
years.  All of these objects show phenomenologically similar behaviors in 
shear flows --- either a tank-treading rotation with a stationary 
shape and a finite inclination angle $\theta>0$ with respect to the flow 
direction, or an unsteady tumbling motion.  However, the qualitative and 
quantitative behavior of the various objects can be very different, 
because the energies governing their shapes and thermal fluctuations are
very different.  In the case of fluid vesicles, which we want to investigate
in this letter, the membrane conformations are determined by the 
curvature elasticity together with the constraints of membrane 
incompressibility and constant internal volume.
 
In simple shear flows, with flow velocity ${\bf v}=\dot\gamma y {\bf e}_x$, 
a transition occurs from tank-treading to tumbling with an increasing viscosity 
of the internal fluid $\eta_{\rm {in}}$~\cite{kant06,beau04,kell82} 
or membrane viscosity $\eta_{\rm {mb}}$~\cite{nogu04,nogu05}.
This transition 
is described very well by the theory of Keller and Skalak 
(K-S)~\cite{kell82}, which assumes a fixed ellipsoidal vesicle shape.
However, shear forces can be large enough to induce shape transformations of
fluid vesicles, for example from discocyte to prolate at small membrane 
viscosity and small viscosity contrast between inside and outside, or 
from prolate to discocyte at larger membrane viscosities \cite{nogu04,nogu05}.
In this case, it is essential to take the membrane deformability into 
account.

Recently, Kantsler and Steinberg~\cite{kant06} reported the first observation
of a new type of vesicle dynamics in shear flow, which is characterized by  
small oscillations of the inclination angle $\theta$ and the deformation,
where $-\theta_0\lesssim \theta(t) \lesssim \theta_0$ with $\theta_0/\pi <1$ 
and time average $\langle\theta\rangle\simeq 0$. 
The vesicles were found to transit from tumbling to this oscillatory motion 
with increasing shear rate $\dot\gamma$. It is worth mentioning that such
an oscillatory motion was also observed in our previous simulations 
(see Fig. 6 in Ref.~\onlinecite{nogu05}), but not further analyzed.
Simultaneously with the experiment, Misbah~\cite{misb06} predicted 
a ``vacillating breathing" mode for quasi-spherical fluid vesicles. 
This mode exhibits very similar dynamical behavior as seen experimentally; 
however, it ``coexists" with the tumbling mode, and its orbit depends on 
the initial deformation i.e. it is not a limit cycle. Furthermore, the shear
rate only enters the theory as the basic time scale, and therefore cannot
induce any shape transitions.
Hence it does not explain the tumbling-to-oscillatory transition seen in 
the experiments \cite{kant06}.

In this letter, we study the oscillatory dynamics of fluid vesicles, which
we will refer to as the ``swinging mode'', by mesoscale hydrodynamics 
simulations and a simplified non-linear theoretical model. The main questions
we want to address are: How does the bending energy affect the 
dynamics? Can transitions between modes be induced by varying the shear rate?
What happens beyond the quasi-spherical limit, which is the typical 
experimental situation? What is the effect of the membrane viscosity?
Is the swinging mode stable when thermal membrane 
undulations are taken into account? We will show that  
the experiments of Ref.~\onlinecite{kant06} can be understood very well on 
the basis of our theory. Furthermore, we will present a complete phase 
diagram of vesicle motion as a function of shear rate and viscosity.

It is worth mentioning that elastic capsules~\cite{chan93} 
and red blood cells~\cite{vall06} can also exhibit a swinging mode; however,
in this case,  the angle $\theta(t)$ is always positive during the oscillation.
Very recently, this dynamics was explained by the K-S theory with an 
addition of an energy barrier for the tank-treading rotation caused by the 
membrane shear elasticity~\cite{skot06}. Therefore, this mechanism cannot
be employed to explain the swinging mode of fluid vesicles. 

The vesicle dynamics is described by several dimensionless quantities.
For a vesicle of volume $V$ and surface area $S$,  
the reduced volume $V^*$ and the excess area $\Delta_{\rm S}$ are 
defined by
$V^*= (R_{\rm V}/R_{\rm S})^3= (1+\Delta_{\rm S}/4\pi)^{-3/2}$ and
$\Delta_{\rm S}=S/R_{\rm V}^2 -4\pi$,
where $R_{\rm V}= (3V/4\pi)^{1/3}$ and $R_{\rm S}=(S/4\pi)^{1/2}$.
The relative viscosity of the inside fluid and membrane are 
$\eta_{\rm {in}}^*=\eta_{\rm {in}}/\eta_{\rm 0}$
and $\eta_{\rm {mb}}^*=\eta_{\rm {mb}}/\eta_{\rm 0}R_{\rm S}$, where 
$\eta_{\rm 0}$ is the viscosity of the outside fluid. 
The shape relaxation time of vesicles with bending rigidity $\kappa$ is given
by $\tau=\eta_{\rm 0}R_{\rm S}^3/\kappa$ (for $\eta_{\rm {in}}^*=1$). This
time is used to define a reduced shear rate $\dot\gamma^*=\dot\gamma \tau$.

The hydrodynamics of fluid vesicle can be studied very well by hybrid 
simulations of a dynamically-triangulated membrane model~\cite{gg:gomp97f} and 
a particle-based mesoscale solvent, multi-particle collision dynamics
\cite{male99}.  A detailed description of this method can be 
found in Ref.~\onlinecite{nogu05}.

\begin{figure}
\includegraphics{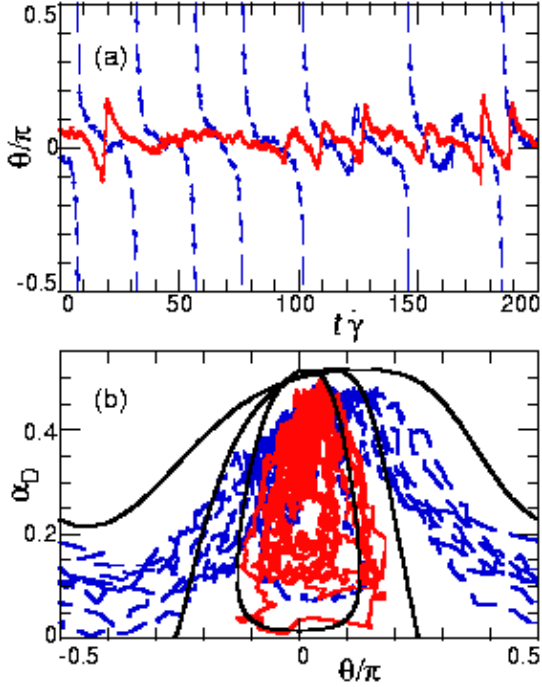}
\caption{ \label{fig:rq}
(Color online)
Temporal evolution of $\alpha_{\rm D}$ and inclination angle 
$\theta$, for $V^*=0.78$ and $\eta_{\rm {mb}}^*=2.9$.
The solid (red) and dashed (blue) lines represent simulation data
for $\dot\gamma^*=3.68$ and $0.92$ ($\kappa/k_{\rm B}T=10$ and $40$ with 
$\dot\gamma \eta_{\rm 0}R_{\rm S}^3/k_{\rm B}T=36.8$), respectively.
The solid lines in (b) are obtained from Eqs.~(\ref{eq:ald}),  
(\ref{eq:thetks}) without thermal noise for $\dot\gamma^*=1.8$, $3$ and 
$10$ (from top to bottom).
}
\end{figure}

\begin{figure}
\includegraphics{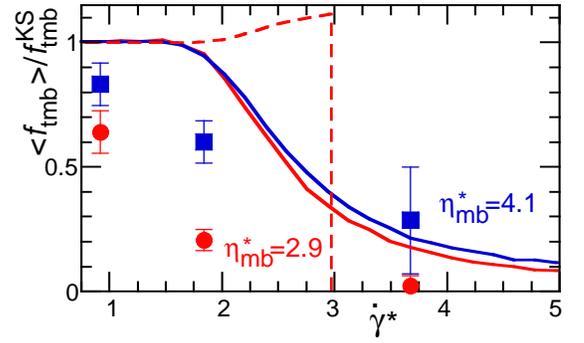}
\caption{ \label{fig:ftum}
(Color online)
$\dot\gamma$ dependence of
the tumbling frequency $f_{\rm {tum}}$ normalized by 
the frequency $f_{\rm {tum}}^{\rm {KS}}=\dot\gamma\sqrt{1-B^2}/2\pi$ of 
K-S theory at $V^*=0.78$.
The bending rigidity $\kappa$ is varied to change $\dot\gamma^*$
with $\dot\gamma \eta_{\rm 0}R_{\rm S}^3/k_{\rm B}T=36.8$.
The symbols represent simulation data for $\eta_{\rm {mb}}^*=2.9$ ($\bullet$) 
and $4.1$ ($\blacksquare$). The solid and dashed lines are obtained from 
Eqs.~(\ref{eq:ald}), (\ref{eq:thetks}) with and without thermal noise, 
respectively.
}
\end{figure}

We have simulated a prolate vesicle with a viscous membrane at $V^*=0.78$ 
and $\eta_{\rm {in}}^*=1$. 
Figure~\ref{fig:rq} shows the time development of the vesicle shape and 
$\theta$.  The shape parameter is $\alpha_{\rm D}=(L_1-L_2)/(L_1+L_2)$, 
where $L_1$ and $L_2$ are the maximum lengths in the direction
of the eigenvectors of the gyration tensor in the vorticity ($x,y$) plane.
The vesicle is found to exhibit a tumbling motion for $\dot\gamma^*=0.92$,
but a swinging motion for $\dot\gamma^*=3.68$.  As $\dot\gamma^*$ increases,
the tumbling frequency $f_{\rm {tmb}}$ decreases, see Fig.~\ref{fig:ftum}, 
while the frequency of the swinging motion increases.
The shear rate $\dot\gamma^*$ at the tumbling-to-swinging
transition is found to increase with increasing membrane viscosity 
$\eta_{\rm {mb}}^*$. These results are consistent with the experiments
of Ref.~\onlinecite{kant06}.

In order to obtain a better understanding of the vesicle dynamics, we
now derive an approximate theoretical model. First, we follow 
Refs.~\onlinecite{seif99,misb06} and employ the Stokes approximation 
and perturbation theory for quasi-spherical vesicles.
The vesicle shape is expanded in spherical harmonics $Y_{l,m}$ as 
$R= R_{\rm V}(1+\sum_{l,m}u_{l,m}Y_{l,m})$. 
The free energy of the membrane with bending rigidity $\kappa$ and surface 
tension $\sigma$ is $F =  \int dS \{\sigma + (\kappa/2)(C_1+C_2)^2 \}$,
where $C_1$ and $C_2$ are the 
principal curvatures at each point of the membrane.
With the harmonic approximation, it is given by
$F = (\kappa/2)\sum_{l,m} E_l |u_{l,m}|^2 + O({u_{l,m}}^3)$,
where $E_l=(l+2)(l-1)\{l(l+1)+ \sigma {R_{\rm V}}^2/\kappa\}$.
The flow fields inside and outside of the vesicle are described by the Lamb 
solution~\cite{seif99}, with no-slip boundary conditions on the membrane.
The flow stress on the membrane is balanced with the elastic forces due 
to bending and tension. This implies that the undulation amplitudes 
$u_{l,m}$ are determined by
\begin{equation}
\label{eq:lamb}
\frac{\partial u_{l,m}}{\partial t} = \frac{i \dot\gamma m}{2} u_{l,m}- 
\frac{\kappa\Gamma_l E_l}{2\eta_0 R_{\rm V}^3}{u_{l,m}}
\mp i h\dot\gamma \delta_{l,2}\delta_{m,\pm 2},
\end{equation}
where $h=60\sqrt{2\pi/15}/(32+23\eta_{\rm {in}}^*)$ and
$\Gamma_l = l(l+1)/\{(l+2)(2l^2-l+2)+(l-1)(2l^2+5l+5)\eta_{\rm {in}}^*\}$.
The detailed derivation of Eq.~(\ref{eq:lamb}) for
$\eta_{\rm {in}}^*=1$ is described in Ref.~\onlinecite{seif99}.

In this letter, we focus on vesicles, which are symmetric with 
respect to the vorticity plane through their center.
Hence, we only consider $u_{2,\pm 2},u_{2,0}$, and
decompose $u_{2,\pm 2}$ into amplitude and phase, 
$u_{2,\pm 2}=r \exp(\mp 2i \theta)$,
where $\theta$ corresponds to the inclination angle \cite{misb06}.
Since the curvature energy does not contribute in this 
case, compare Ref.~\onlinecite{misb06}, we have to go beyond the harmonic 
approximation.
Therefore, we replace the force $\kappa E_l r$ in Eq.~(\ref{eq:lamb}) 
by $\partial F/\partial r$. This implies 
\begin{eqnarray}
\label{eq:r2f}
\frac{\partial r}{\partial t} &=&  -\frac{\Gamma_2}{2\eta_0 R_{\rm V}^3} 
   (\frac{\partial F}{\partial r} + 2\lambda r) + \dot\gamma h\sin(2\theta),\\
\label{eq:thetlam}
\frac{\partial \theta}{\partial t} &=& \frac{\dot\gamma}{2}
                \{ -1 + \frac{h}{r}\cos(2\theta)\},\\
\label{eq:u20}
\frac{\partial u_{2,0}}{\partial t} &=&  -\frac{\Gamma_2}{\eta_0 R_{\rm V}^3} 
         (\frac{\partial F}{\partial u_{2,0}} + \lambda u_{2,0}).
\end{eqnarray}
A Lagrange multiplier $\lambda\equiv \sigma R_V^2$ is employed to
satisfy the area constraint \cite{miln87}; it is determined by
$d{\Delta}_{\rm S}/dt=0$, which implies 
$\Delta_{\rm S}= 2 {u_{2,0}}^2 + 4 r^2$. Thus, Eq.~(\ref{eq:r2f}) 
becomes
\begin{equation}
\label{eq:r2ds}
\frac{d r}{dt} = \left\{1-\frac{4r^2}{\Delta_{\rm S}}\right\}
       \left\{ -\frac{\Gamma_2}{2\eta_0 R_{\rm V}^3} 
       \frac{\partial F}{\partial r}\Bigl|_{\Delta_{\rm S}} 
                      + \dot\gamma h\sin(2\theta)\right\},
\end{equation}
where $\partial F/\partial r|_{\Delta_{\rm S}} = 
        \partial F/\partial r-(2r/u_{2,0})\partial F/\partial u_{2,0}$.
In the harmonic approximation of $F$,
the first term in Eq.~(\ref{eq:r2ds}) disappears because of 
$\partial F/\partial r|_{\Delta_{\rm S}}=0$, so that 
we recover the description of Ref.~\onlinecite{misb06}.
The prolate shape appears as an energy minimum,
when higher-order terms in the free energy $F$ are taken into 
account~\cite{miln87}.

\begin{figure}
\includegraphics{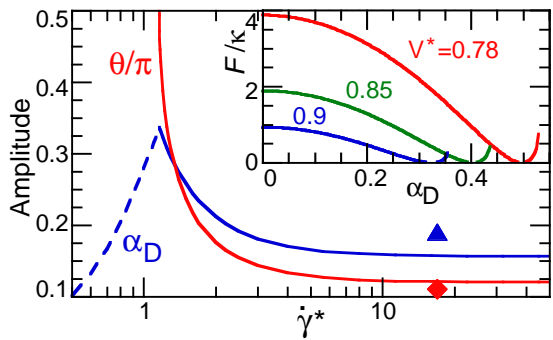}
\caption{ \label{fig:cv}
(Color online)
$\dot\gamma$ dependence of the amplitudes of the oscillations of 
$\alpha_{\rm D}$ and $\theta$ 
for $V^*=0.9$, $\eta_{\rm {in}}^*=6$, and $\eta_{\rm {mb}}^*=0$.
The dashed and solid lines represent the tumbling and swinging states, 
respectively, given by Eqs.~(\ref{eq:ald}), (\ref{eq:thetks})
without thermal noise.  Symbols represent  
estimates from the experimental data~\cite{kant06} for
$\alpha_{\rm D}$ ($\blacktriangle$) and $\theta$ ($\blacklozenge$).
The inset shows the curvature energy $F(\alpha_{\rm D})$ of an ellipsoid
for the reduced volumes $V^* = 0.78$, $0.85$, and $0.9$.
}
\end{figure}

Since an expansion of the vesicle shape in spherical harmonics
is difficult experimentally, more easily accessible measures of the 
deformation are desirable. Furthermore, a description is needed which
goes beyond the quasi-spherical limit. Thus, instead of $r$, we employ 
the shape parameter $\alpha_{\rm D}=(L_1-L_2)/(L_1+L_2)$,
which is easily measurable by microscopy.
Since $\alpha_{\rm D}= (\sqrt{15/2\pi}/2) r + O(r^2)$,
Eq.~(\ref{eq:r2ds}) implies
\begin{equation}
\label{eq:ald}
\frac{d \alpha_{\rm D}}{dt} = \left\{1-\left(\frac{\alpha_{\rm D}}
                   {\alpha_{\rm D}^{\rm {max}}}  \right)^2\right\}
    \left\{ -\frac{A_0}{\tau\kappa V^*} 
       \frac{\partial F}{\partial \alpha_{\rm D}} 
                + \dot\gamma A_1\sin(2\theta)\right\},
\end{equation}
where $A_0= 15\Gamma_2/16\pi= 45/8\pi(32+23\eta_{\rm {in}}^*)$ and 
$A_1= h\sqrt{15/2\pi}/2= 30/(32+23\eta_{\rm {in}}^*)$.
Since an accurate evaluation of the free energy $F$ is very important, 
we calculate it numerically for ellipsoidal vesicles with 
$(x_1/a_1)^2 + (x_2/a_2)^2 + (x_3/a_3)^2=1$, see the inset in Fig.~\ref{fig:cv}.
The prolate ($a_1>a_2=a_3$) and oblate ($a_1=a_2>a_3$) shapes are energy 
minima and maxima, respectively,
and $\partial F/\partial \alpha_{\rm D}$ diverges in the limit of maximum
extension, $\alpha_{\rm D} \to \alpha_{\rm D}^{\rm {max}}(V^*)$.
Eq.~(\ref{eq:ald}) has the same form as the simplified model studied previously 
\cite{nogu04,nogu05}, but now has {\em no adjustable parameters}.

In a final step, in order to obtain a reliable description also for large
excess areas, we replace Eq.~(\ref{eq:thetlam}) by the equation of K-S 
theory~\cite{kell82,tran84}, which reads
\begin{eqnarray}
\label{eq:thetks}
\frac{d \theta}{dt} &=& \frac{\dot\gamma}{2}\{-1+B\cos(2\theta)\},\\
\label{eq:KS-B} \nonumber
B &=& f_0\left\{f_1+ \frac{f_1^{-1}}
      {1+f_2(\eta_{\rm {in}}^* -1) 
                  + f_2f_3 \eta_{\rm {mb}}^*}\right\},
\end{eqnarray}
where the factors $f_0$, $f_1$, $f_2$, and $f_3$ are
the functions of the ellipsoidal shape ($a_2/a_1$, $a_3/a_1$),
and are given in Appendix B of Ref.~{\onlinecite{nogu05}}.
The K-S theory in general shows very good agreement with simulation 
results~\cite{krau96,nogu05}.
When thermal fluctuations are taken into account,
Gaussian white noises $g_{\alpha}(t)$ and $g_{\theta}(t)$
are added to Eqs.~(\ref{eq:ald}) and (\ref{eq:thetks}), respectively,
which obey the fluctuation-dissipation theorem, so that
$\langle g_i(t)\rangle = 0$,
$\langle g_i(t) g_j(t')\rangle =
      2 k_{\text B}T/\zeta_i \delta_{ij}\delta(t-t')$, 
where $i,j \in \{\alpha_{\rm D},\theta\}$.
The friction coefficients are $\zeta_{\alpha}=
    \tau\kappa V^*/A_0\{1-(\alpha_{\rm D}/\alpha_{\rm D}^{\rm {max}})^2\}$ 
and the rotational friction coefficient of a sphere,  
$\zeta_{\theta}=8\pi \eta_{\rm {0}}R_{\rm S}^3$.
We numerically integrate Eqs.~(\ref{eq:ald}) and (\ref{eq:thetks})
with or without thermal noise 
using the second or fourth-order Runge-Kutta method, respectively.
When $\alpha_{\rm D}(t)$ becomes zero, we set 
($\alpha_{\rm D}$,$\theta$)=($0,\pi/4$), since $\theta=\pi/4$ is 
the inclination angle in the spherical limit.

Eqs.~(\ref{eq:ald}) and (\ref{eq:thetks}) reproduce the simulation results 
semi-quantitatively (see Figs.~\ref{fig:rq} and \ref{fig:ftum}).
This is a very good agreement, given the fact that our model certainly
does not systematically take into account all higher order terms, and
furthermore has no adjustable parameters.
At small $\dot\gamma^*$, in the tumbling phase,
$\theta$ rotates with $\alpha_{\rm D}$-oscillation of small amplitude.
This $\alpha_{\rm D}$ amplitude becomes larger at larger $\dot\gamma^*$,
see Fig.~\ref{fig:cv}.  Then, when $\alpha_{\rm D}(t)$ reaches zero,
$\theta$ jumps to $\pi/4$, see Fig.~\ref{fig:rq}.
This type of $\theta$ oscillations with a jump is reminiscent  
of the behavior predicted for
very viscous liquid drops within perturbation theory~\cite{bart85}.
Finally, at even larger $\dot\gamma^*$,
$\alpha_{\rm D}$ and $\theta$ exhibit oscillations without jumps. 

The physical mechanism of swinging can be understood
on the basis of Eqs.~(\ref{eq:ald}), (\ref{eq:thetks}) as follows.
At finite $\kappa$, the shear force elongates the vesicle 
($\alpha_{\rm D}$ increases) for $0<\theta<\pi/2$,
but compresses it ($\alpha_{\rm D}$ decreases) for $-\pi/2<\theta<0$,
since the $\sin(2\theta)$-term in Eq.~(\ref{eq:ald}) changes sign.
Thus, the swinging motion is caused 
by a shape deformation, where $B$ in Eq.~(\ref{eq:thetks})
crosses the tank-treading-to-tumbling threshold periodically.
First, a prolate vesicle starts tumbling because $B<1$, 
$\alpha_{\rm D}$ decreases when $\theta<0$, which implies that $B$ increases;
then $\theta$ starts to increase again because $B>1$ at small 
$\alpha_{\rm D}$; finally $\alpha_{\rm D}$ increases when $\theta>0$.
In the swinging phase, the amplitudes of $\alpha_{\rm D}$ and $\theta$ 
decrease and saturate to finite values with increasing $\dot\gamma^*$, 
see Fig.~\ref{fig:cv}.

A linear stability analysis of the fixed points of Eqs.~(\ref{eq:ald}), 
(\ref{eq:thetks}) shows that the tank-treading-to-tumbling transition 
always occurs as a saddle-node bifurcation at $\theta\simeq 0$.
The tank-treading-to-swinging transition is a saddle-node bifurcation for 
small $\dot\gamma^*$, but becomes a Hopf bifurcation at $\theta<0$ for larger 
$\dot\gamma^*$.  Near the boundary of these two bifurcations,
a second range of stable fixed points appears between the saddle and 
unstable fixed points at $\theta<0$.
However, these stable points at $\theta<0$ will likely disappear
when the shape degrees of freedom in the vorticity direction 
are taken into account. 

\begin{figure}
\includegraphics{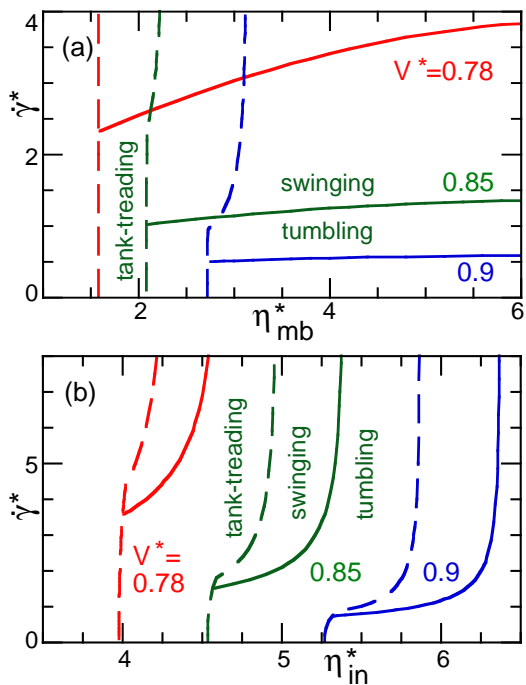}
\caption{ \label{fig:phase}
(Color online)
Dynamical phase diagrams as a function of (a) $\eta_{\rm {mb}}^*$ for 
$\eta_{\rm {in}}^*=1$, and (b) $\eta_{\rm {in}}^*$ for $\eta_{\rm {mb}}^*=0$,
for various reduced volumes $V^*$, obtained from Eqs.~(\ref{eq:ald}), 
(\ref{eq:thetks}) without thermal noise.
The tank-treading phase is located on the left-hand-side of the dashed 
lines. The solid lines represent the tumbling-to-swinging transitions.
}
\end{figure}

Fig.~\ref{fig:phase} shows the full phase diagrams, both for variations of
the internal and the membrane viscosities.
With increasing reduced volume $V^*$ the smallest shear rate 
$\dot\gamma^*_{\rm {os}}$, at which swinging can be observed,  
decreases, since the dependence of the energy $F$ on $\alpha_{\rm D}$
becomes more shallow, compare Fig.~\ref{fig:cv}. Also, the region of
stability of swinging shifts to higher viscosity ($\eta_{\rm in}$ or 
$\eta_{\rm mb}$) with increasing $V^*$. On the other hand, for fixed $V^*$,
$\dot\gamma^*_{\rm {os}}$ is a monotonically increasing function of
$\eta_{\rm {mb}}^*$ or $\eta_{\rm {in}}^*$. This increase is found to be
much less pronounced for $\eta_{\rm {mb}}^*$. The reason is that in our
derivation of Eq.~(\ref{eq:lamb}), we have neglect the $\eta_{\rm {mb}}^*$ 
dependence. In analogy with Eq.~(\ref{eq:thetks}), it is plausible to expect 
that in a more detailed calculation, a linear combination of $\eta_{\rm {mb}}^*$ 
and $\eta_{\rm {in}}^*$ will appear in the both $h$ and $\Gamma_l$.  
If this $\eta_{\rm {mb}}^*$ dependence were taken into account,
the $\dot\gamma^*_{\rm {os}}$ lines in Fig.~\ref{fig:phase}(a) will curve 
upwards more strongly; this is consistent with
the larger $\eta_{\rm {mb}}^*$ dependence observed in the simulations in 
Fig.~\ref{fig:ftum}.

In the experiments of Ref.~\onlinecite{kant06}, the swinging (tumbling) motion 
was observed for $\dot\gamma^*\simeq 17$ ($4.5$), $\eta_{\rm {in}}^*=6$ ($8.4$), 
and  $V^*\simeq 0.9$, and a very small membrane viscosity 
$\eta_{\rm {mb}}^* \sim 0.1$ (calculated from 
$\eta_{\rm {mb}} \sim 10^{-9}$ Ns/m of Ref.~\onlinecite{dimo99}).
It is mentioned that the swinging motion is seen in particular in 
the close vicinity to the tank-treading-to-tumbling transition.
Thus, the experimental data confirm our predicted phase diagram, 
Fig.~\ref{fig:phase}, very well.
Furthermore, the oscillation amplitudes also show good agreement, 
see Fig.~\ref{fig:cv}.

In summary, we have studied the oscillatory motion of fluid vesicles in 
simple shear flow.
We have developed a simplified model for the ellipsoidal fluid vesicle,
which explains the simulation and experimental results very well.
This model could be extended in the future to study the coupling of  
different types of 
oscillation mechanisms, like membrane shear elasticity and viscosity, 
in elastic capsules or red blood cells.

\begin{acknowledgments}
We thank R. Finken (Stuttgart) for his help to rederive Eq.~(\ref{eq:lamb}).
GG acknowledges support of this work through the DFG priority program 
``Nano- and Microfluidics''.

\end{acknowledgments}

\bibliographystyle{apsrev}

\end{document}